\documentclass [aps, pra, twocolumn, twoside, 10pt, amsfonts, amsmath, nofootinbib] {revtex4-1}

\usepackage [inline] {asymptote}
\usepackage {asyfig}
\usepackage {bigints}
\usepackage {bm}
\usepackage {colortbl}
\usepackage {mathtools}
\usepackage {microtype}
\usepackage {physics}
\usepackage {relsize}
\usepackage [hyperfootnotes=false] {hyperref}

\pdfoutput=1

\usepackage {perpage}
\MakePerPage {footnote}

\makeatletter
\newcommand{\footnoteref}[1]{%
\ltx@ifpackageloaded{hyperref}{
  \ifHy@hyperfootnotes
    \hbox{\hyperref[#1]{%
            \@textsuperscript {\normalfont \ref*{#1}}}}%
  \else
    \hbox{\@textsuperscript {\normalfont \ref*{#1}}}%
  \fi%
}{
    \hbox{\@textsuperscript {\normalfont \ref{#1}}}%
 }%
}
\makeatother

\setcitestyle {super, open={}, close={}}

\newcommand {\dt} {\dd {\tau}}

\newcommand {\HH} {\mathbf H}
\newcommand {\HHp} {\mathbf {H'}}
\newcommand {\HHz} {\mathbf {H_0}}
\newcommand {\XX} {\mathbf X}
\newcommand {\JJ} {\mathbf J}
\newcommand {\DDelta} {\mathbf \Delta}
\newcommand {\UU} {\mathbf U}
\newcommand {\EEta} {{\bm \eta}}
\newcommand {\EEtah} {\bm {\eta_\mathit h}}
\newcommand {\EEtahp} {\bm {\eta_\mathit h'}}
\newcommand {\EEtaz} {\bm {\eta_0}}
\newcommand {\CCdot} {\bm \cdot}
\newcommand {\RR} {\mathbf R}
\newcommand {\TTheta} {\mathbf \Theta}
\newcommand {\ZZero} {\mathbf 0}
\newcommand {\EE} {\mathbf E}
\newcommand {\ZZeta} {\bm \zeta}
\newcommand {\ZZetap} {\bm {\zeta'}}
\newcommand {\ZZetaz} {\bm {\zeta_0}}
\newcommand {\ZZetazp} {\bm {\zeta_0'}}
\newcommand {\ZZetazpp} {\bm {\zeta_0''}}

\newcommand {\OO} {\mathbf O}
\newcommand {\AAA} {\mathbf A}
\newcommand {\BB} {\mathbf B}
\newcommand {\SSigma} {\mathbf \Sigma}
\newcommand {\ssigma}[1]{\bm {\sigma_{#1}}}
\newcommand {\XXi} {\mathbf \Xi}

\DeclareMathOperator {\aatan} {tan^{-1}}
\DeclareMathOperator {\aasin} {sin^{-1}}
\DeclareMathOperator {\atantwo} {atan2}
\DeclareMathOperator {\sgn} {sgn}
\DeclareMathOperator {\tantan} {tan^2}
\DeclareMathOperator {\gd} {gd}

\newcommand {\F} [2] {\DeclareMathOperator {#1} {\mathnormal {#2 \!}}}
\newcommand {\FF} [2] {\DeclareMathOperator {#1} {\mathbf {#2 \!}}}

\F \HHF \HH
\F \xF x
\F \jF j

\F \epsilonF \epsilon
\F \yF y

\F \EEtaF \EEta
\F \thetaF \theta

\FF \EEtahF \EEtah
\FF \EEtazF \EEtaz
\FF \EEtahpF \EEtahp

\F \zetaF \zeta
\F \zetapF {\zeta'}
\F \zetappF {\zeta''}

\F \ZZetaF \ZZeta
\F \ZZetazF \ZZetaz
\F \ZZetapF \ZZetap

\F \zetaroF {\zeta_{r_0}}

\F \oF {\mathfrak o}

\F \sigmaF \sigma

\F \JJFab {\JJ_{ab}}

\begin{asydef}
  import graph;
\end{asydef}

\begin {document}

\title {Stable Unitary Integrators for the Numerical Implementation of Continuous Unitary Transformations}

\author {Samuel \surname {Savitz}}\email {Sam@Savitz.org}
\author {Gil \surname {Refael}}\email {Refael@Caltech.edu}
\affiliation {Institute of Quantum Information and Matter, Department of Physics, California Institute of Technology, Pasadena,~California~91125,~USA}

\date {July 11\textsuperscript {th}, 2017}

\begin {abstract}
  The technique of continuous unitary transformations has recently been used to provide physical insight into a diverse array of quantum mechanical systems.  However, the question of how to best numerically implement the flow equations has received little attention.  The most immediately apparent approach, using standard Runge--Kutta numerical integration algorithms, suffers from both severe inefficiency due to stiffness and the loss of unitarity.  After reviewing the formalism of continuous unitary transformations and Wegner's original choice for the infinitesimal generator of the flow, we present a number of approaches to resolving these issues including a choice of generator which induces what we call the ``uniform tangent decay flow'' and three numerical integrators specifically designed to perform continuous unitary transformations efficiently while preserving the unitarity of flow.  We conclude by applying one of the flow algorithms to a simple calculation that visually demonstrates the many-body localization transition.\\\\
  DOI: \href {https://doi.org/10.1103/PhysRevB.96.115129} {10.1103/PhysRevB.96.115129}
\end {abstract}

\maketitle

\section {Introduction} \label {Introduction}

While the technique of continuous unitary transformations (CUTs) was already known to contemporary mathematicians,\cite {DeiftEtAl83} it was introduced to the physics community roughly simultaneously by Franz Wegner in the language of ``flow equations''\cite {Wegner94} and by Kenneth Wilson and Stanis{\l}aw G{\l}azek who recast it as a renormalization group\cite {GlazekWilson93, GlazekWilson94}.  Thus, the terms ``Wegner flow'' or ``Wegner--Wilson flow'' (WWF) are established in the condensed matter literature as referring to a particular kind of CUT.  The applications and mathematical properties of CUTs have been reviewed by Monthus\cite {Monthus16} and Bartlett\cite {Bartlett03}.  They have also been analyzed in terms of geodesic flows.\cite{IttoAbe12}

Interest in CUTs has increased recently due to the discovery that they may provide physical insights when applied to important many-body problems such as the poorly understood phenomenon of many-body localization (MBL).\cite {HeidbrinkUhrig02, KehreinMielke94}  The study of MBL has exploded since Basko, Aleiner, and Altshuler published their seminal diagrammatic analysis of the phenomenon in 2006\cite {BaskoEtAl06}.  Our understanding of MBL is deeply connected to the concepts of ergodicity and thermalization in quantum systems and has been reviewed by Nandikishore and Huse\cite {NandikishoreHuse15}.  Recent results of this approach include using the WWF to find $\ell$-bits representations of an MBL Hamiltonian\cite {PekkerEtAl16, RademakerOrtuno16} and to identify an Anderson transition in the power-law banded random matrix model\cite {QuitoEtAl16}.  While we will not address MBL directly until section~\ref {MBL}, we have used it as a source of interesting initial Hamiltonians in figures~\ref {NovelPlot} and~\ref {OrderPlot}.

Despite this growing interest, little has been said about how best to numerically implement these flows \emph {in silico}.  In Kehrein's monograph on the application of CUTs to many-body problems\cite {Kehrein06}, the standard Runge--Kutta family of adaptive numerical integrators, including the popular Dormand--Prince method\cite {DormandPrince80}, is described as a ``good algorithm'' for this purpose.  Nevertheless, we have identified a number of ways in which it can be dramatically improved.  In particular, our methods eliminate a severe inefficiency near the end of the flow, once many of the off-diagonal elements have been almost entirely eliminated.  A second flaw exhibited by the current flow algorithms which we resolve is the violation of the unitarity of the transformation in a step size-dependent manner, leading to subsequent errors in the final eigenvalues.

After first reviewing the formalism of CUTs, we will proceed to explain the origins of these issues in section~\ref {Issues}.  In section~\ref {NovelFlow}, we will propose a choice of infinitesimal generator that leads to what we call the ``uniform tangent decay flow'' and which elegantly resolves the efficiency problem without having to change integrators.  Then, in section~\ref{UnitaryIntegrator}, we will solve the efficiency and unitarity issues, regardless of the choice of generator, by developing two stable, geometric\cite {McLachlanQuispel06, Hairer02} integrators designed specifically for CUTs.  One is accurate to first-order, and the other to third-order.  In section~\ref {Trotter}, we will present an entirely different approach to resolving these problems that was inspired by quantized state simulators\cite {CellierEtAl08, MigoniEtAl11} and the Trotter decomposition\cite {Trotter59}.  Finally, we will demonstrate a simple way in which CUTs can provide insight and help visualize the MBL transition before giving some concluding thoughts regarding the interpretation of, and potential future directions for, the study of CUTs.

\section {Formalism of Continuous Unitary Transformations} \label {Formalism}

\subsection {Notation}

We begin with a Hamiltonian $\HH$ represented by an $n \times n$ Hermitian matrix.  For the sake of simplicity and clarity, its elements, $\HH_{ab}$, will be restricted to be real, although nothing prevents what follows from being extended to complex matrices.  We will refer to the diagonal elements by $D_a = \HH_{aa}$, and the off-diagonal elements by $\JJ_{ab} = \JJ_{ba} = \HH_{ab}$.  Furthermore, we will use $\DDelta_{ab} = D_a - D_b$ to represent the current energy spacing between any two distinct states $\ket a$ and $\ket b$, ignoring off-diagonal elements.  $\XX_{ab} = -\XX_{ba} = \DDelta_{ab}/2$ will be half this value.

A generic unitary transformation applied to $\HH$ can be represented by $\HHp = \UU \HH \UU^{-1}$ where $\UU$ is a unitary matrix ($\UU^{-1} = \UU^{\dag}$).  This can also be written as
\begin {equation}
  \HHp = e^\EEta \HH e^{-\EEta}, \label {Exponential}
\end {equation}
where $\EEta$ is the anti-Hermitian ($\EEta^\dag = -\EEta$) matrix logarithm of $\UU$, which we will also assume to have only real elements\footnotetext {\label{FNReal}Note that this implies that $\HH$ is actually symmetric, $\UU$ is orthogonal, and $\EEta$ is antisymmetric.  Again, extending this work to complex matrices should not be difficult.}\footnoteref {FNReal}.

\subsection {The Flow Equations}

In the limit of a small unitary rotation, we can replace $\EEta$ with $\dt \EEta$.  $\UU$ converges to the identity matrix as the infinitesimal $\dt$ approaches zero.  Truncating the Taylor series expansion of the exponential,
\begin {equation}
  \UU = e^{\dt \EEta} = \mathlarger{\mathlarger{\sum}}_{k = 0}^\infty \frac {\dt^k} {k!} \EEta^k, \label {ExponentialSeries}
\end {equation}
after the second term gives $\UU \approx 1 + \dt \EEta$, and $\UU^{-1} \approx 1 - \dt \EEta$.  Therefore, to first-order,
\begin {equation}
  \HHp \approx (1 + \dt \EEta) \HH (1 - \dt \EEta) \approx \HH + \dt \comm {\EEta} \HH.
\end {equation}

Treating $\tau$ as a fictitious ``flow-time'' coordinate, this can be recast as the first-order, nonlinear differential equation
\begin {equation}
  \dv{\HH}{\tau} = \dot \HH = \comm {\EEta} \HH. \label {Flow}
\end {equation}
The diagonalizing unitary evolves according to $\dot \UU = \EEta \UU$.  These differential equations are known as the ``flow equations''.  While $\UU$ begins as the identity matrix, $\HH$ is initialized to whatever starting Hamiltonian one desires: $\HHF {(\tau = 0)} = \HHz$.  Written in terms of the elements of $\HH$, the flow equations are
\begin {samepage}\begin {subequations} \label {FFlow} \begin {alignat} {2}
  \dot D_a = \, && 2 & \sum_{\mathmakebox[0pt][l]{c \ne a}\phantom{c \ne a, b}} \EEta_{ac} \JJ_{ca}, \label {DFlow} \\
  \dot \XX_{ab} = \, && 2 \EEta_{ab} \JJ_{ab} + & \sum_{c \ne a, b} \left[ \EEta_{ac} \JJ_{ca} - \EEta_{bc} \JJ_{cb} \right], \label {XFlow} \\
\intertext {and}
  \dot \JJ_{ab} = \, && -2 \EEta_{ab} \XX_{ab} + & \sum_{c \ne a, b} \left[ \EEta_{bc} \JJ_{ca} + \EEta_{ac} \JJ_{cb} \right]. \label {JFlow}
\end {alignat}\end {subequations}\end {samepage}
Equation~\eqref {XFlow} follows immediately from equation~\eqref {DFlow} and possesses a pleasing symmetry with equation~\eqref {JFlow}.

\subsection {The Two-State Limit} \label {TwoState}

The proper interpretation of the generator $\EEta$ can be elucidated by recasting the flow in angular terms.  In order to clarify the intuitive picture, we will temporarily restrict our attention to the $n = 2$ case of two-by-two matrices.  Equivalently, we can assume that $\EEta$ has only one pair of nonzero elements $\EEta_{ab} = -\EEta_{ba}$.  Either of these restrictions allows us to neglect the summed terms in equations~(\ref {FFlow}bc).  We will refer to this approximation as the ``two-state limit'', and it will be recurrently useful.  We will henceforth sometimes lighten and decapitalize the upper case letters denoting matrices and drop the $ab$ subscripts when they do not enhance clarity.\footnotetext {\label {FNHadamard}Decapitalized matrix equations therefore represent the entry-wise Hadamard analog of their bolded versions.}\footnoteref {FNHadamard}

Without further loss of generality, we can now write
\begin {samepage}\begin {subequations}\begin {alignat} {4}
  \HH & \, = \, & \begin{pmatrix}
    x & j \\
    j & -x
  \end{pmatrix} & \, = \, &&
  j\ssigma {1} + x\ssigma {3}, \\
\intertext {and}
  \EEta & \, = \, & \begin{pmatrix}
    0 & \eta \\
    -\eta & 0
  \end{pmatrix} & \, = \, &&
  i\eta\ssigma {2},
\end {alignat}\end {subequations}\end {samepage}
where the $\ssigma {}$'s denote the standard Pauli matrices.  Conveniently, these matrices will retain this form throughout the flow, so we can understand their evolution as a first-order differential equation for $\xF {(\tau)}$ and $\jF {(\tau)}$.

\subsubsection {The Angular Interpretation}

For each pair of distinct states, we can define an angle $\theta$ to equal $\atantwo{ \left( j, x \right) }$, \emph {i.e.}~the complex phase of the value $z = x + ij$\footnotetext {\label {FNAtanTwo}The function $\atantwo$ has the minor advantage over the usual inverse tangent that there is no $\pm \pi$ ambiguity.}\footnoteref {FNAtanTwo}.  A positive radius $r = \sqrt {x^2 + j^2}$ is also given by its magnitude, $|z|$.  In these polar coordinates, $x = r \cos {\theta}$, and $j = r \sin {\theta}$.  Note that the condition of $\HH$ being successfully diagonalized is, barring degeneracies, equivalent to the requirement that all of the $\theta$'s satisfy $\sin \theta = 0$ by being an integer multiple of $\pi$.

Finally, differentiating $\theta$ according to the chain rule gives
\begin {equation}
  \dot \theta = \frac {x \dot j - j \dot x} {r^2} = -2 \eta. \label {RotationRate}
\end {equation}
Thus, $\EEta_{ab}$ is clearly proportional to the rate of rotation between the $\ket a$- and $\ket b$-axes of the Hamiltonian's basis.  The negative two factor in equation~\eqref {RotationRate} is identical in origin to the phenomenon of angle doubling in spinor homomorphisms.\cite {Steane13}  The complication outside of the two-state limit comes, of course, from the sum terms in equations~(\ref {FFlow}bc), which we are neglecting.  They capture how this rotation between the $\ket a$- and $\ket b$-axes affects the couplings between those two states and any distinct third state, $\ket c$, namely $\JJ_{ac}$ and $\JJ_{bc}$.

\subsection {Choice of Generator}

\begin {table} [htb]
  \centering

  \begin {tabular} {lr|rcl}
    \multicolumn {2} {c|} {Flow} && $\EEta_{ab}$ \\ \hline
    WWF & \cite {Wegner94} & $\delta j$ & $=$ & $r^2 \sin {2 \theta}$ \\
    White & \cite {White02} & $j/\delta$ & $=$ & $\tan {(\theta)}/2$ \\
    Sign & \cite {MorrisEtAl15} & $\sgn {(x)} \, j$ & $=$ & $\sgn {(x)} \, r \sin {\theta}$ \\
    Toda & \cite {Henon74} & $\sgn {(b - a)} \, j$ & $=$ & $\sgn {(b - a)} \, r \sin {\theta}$ \\ \arrayrulecolor [gray] {0.67} \hline \arrayrulecolor [gray] {0}
    Tangent && $\delta j/ \left( x^2 + j^2 \right) $ & $=$ & $\sin {2 \theta}$ \\
  \end {tabular}

  \caption {The best-known flow generators, and our tangent flow, expressed in terms of both the matrix elements and their polar form.}
  \label {Flows}
  \hrule
\end {table}
In order to finish specifying the flow, we must choose a formula for $\EEta$.  Most often $\EEta_{ab}$ is a function of only $\XX_{ab}$, $\JJ_{ab}$, and, in the case of the Toda flow, the sign $\sgn {(b - a)}$.  Defining $\EEta$ purely in terms of the ``current'' value of $\HH$ allows us to think about a given CUT as a time-independent, first-order differential equation flowing over the space of all Hermitian matrices.  Wegner's original choice of generator was $\EEta = \comm {\HH_\mathrm {Diag.}} \HH$, where $\HH_\mathrm {Diag.}$ is the diagonal part of $\HH$.  Expressed in terms of the matrix elements, this specifies that $\eta = \delta j = r^2 \sin {2 \theta}$.  Recall that $\delta = 2x$ is the difference of the two diagonal elements.

As Monthus recounts\cite {Monthus16}, a number of alternative choices of generator have since been presented.  The flows reviewed in her paper are summarized in table~\ref {Flows}.  In conjunction with equation~\eqref {RotationRate}, the polar forms of these generators make it apparent why they must induce evolution towards a diagonalized Hamiltonian.

\subsection {Matrix Metrics and the Convergence Towards Diagonalization}

In order to show more rigorously that these flows ultimately lead to the diagonalization of the Hamiltonians to which they are applied, it will be useful to introduce some metrics which can be applied to the Hamiltonian during the course of the flow:  The Frobenius norm of a matrix, $\norm {\CCdot}_F$, is defined as
\begin {equation}
  \norm {\HH}_F = \sqrt {\sum_{a, b} \HH_{ab}^2} = \sqrt {I_2^D + I_2^J}, \label {Frobenius}
\end {equation}
where\cite {Monthus16}
\begin {samepage}\begin {subequations}\begin {align}
  I_2^D =& \sum_a {D_a^2},
\intertext {and}
  I_2^J =& \sum_{a \ne b} {\JJ_{ab}^2}. \label {I2J}
\end {align}\end {subequations}\end {samepage}
Continuing to decompose terms, we find that
\begin {equation}
  I_2^D = \frac {I_2^\Delta + \left( \Tr \HH \right) ^2 } {n},
\end {equation}
given
\begin {equation}
  I_2^\Delta = \sum_{a < b} \DDelta_{ab}^2 = n^2 \sigma_D^2, \label {I2Delta}
\end {equation}
where
\begin {equation}
  \sigma_D^2 = \frac 1 n \, \mathlarger{\mathlarger{\sum}}_a \left( D_a - \mu \right)^2
\end {equation}
is the population variance of the diagonal elements, and $\mu = \Tr \HH/n$ is their mean.  Both the Frobenius norm and the trace are invariant under unitary transformations.  Therefore, the effect of the flow is to transfer weight directly between $I_2^J$ and $I_2^D$, or equivalently, between $I_2^J$ and $n \sigma^2_D$.

We can now immediately calculate that
\begin {equation}
  \dot {I_2^D} = 4 \sum_{a \ne b} \EEta_{ab} \XX_{ab} \JJ_{ab} = 2 \sum_{a \ne b} \EEta_{ab} \RR_{ab}^2 \sin {2 \TTheta_{ab}}.
\end {equation}
Thus, a choice of generator which ensures that the sign of $\eta$ always matches that of $xj \propto \sin 2\theta$ must cause $I_2^D$ to evolve in a nondecreasing manner.  It is easy to verify that all of the generators in table~\ref {Flows}, with the exception of the Toda flow\footnotetext {\label {FNToda}The Toda flow was rediscovered and introduced to physics by Mielke\cite {Mielke98} and approaches the diagonalized form with the eigenvalues sorted in descending order, where all the $\theta$'s are zero.  Remarkably, it preserves the structure of the banded matrices to which it is applied.\cite {Henon74}  It can be shown to be a discretization of the well-known continuous Korteweg--de Vries nonlinear partial differential equation.}\footnoteref {FNToda}, satisfy this condition.

In fact, the original WWF generator is given by the steepest descent of $I_2^J$:\cite {Bartlett03}
\begin {equation}
  \EEta \propto \left. - \dv{\dot {I_2^J}}{\EEta} \right \vert_{\EEta = \ZZero}. \label {SteepestDescent}
\end {equation}
In other words, each independent component of $\EEta$ is set in proportion to how quickly it causes $I_2^J$ to decrease.  Given a fixed Frobenius norm $\norm {\EEta}_F$, no alternative choice of $\EEta$ leads to a faster decrease than the WWF.

As noted above, the flow towards diagonalization decreases $I_2^J$ and increases $\sigma_D$.  In other words, as $j$ decays towards zero, the corresponding level spacing $\delta$ increases:  The eigenvalues appear to repel each other.  This behavior is related in origin to the Dyson Brownian motion of eigenvalues in random matrix theory.\cite {Dyson62, Mehta04}

One can monitor the progression of a flow towards diagonalization using the metric $\rho$ which we will define in terms of equations~\eqref {I2J} and~\eqref {I2Delta} as
\begin {equation}
  \rho = \sqrt {\frac {2 I_2^J} {I_2^\Delta + 2 I_2^J}} = \sqrt {\frac {2 I_2^J} {n^2 \sigma_D^2 + 2 I_2^J}}. \label {Rho}
\end {equation}
The Hamiltonian is diagonalized when $\rho$ reaches zero.

\subsection {Fixed Points, Viscous Pendula, the Gudermannian function, and the Renormalization Group} \label {Renormalization}

While all stable fixed points of the WWF must be diagonal, an unstable fixed point can have nonzero $\JJ_{ab}$ if the corresponding $\XX_{ab}$ is zero, \emph {i.e.} the relevant diagonal elements are equal.  Our uniform tangent decay flow introduced in section~\ref {NovelFlow} also behaves this way, but the other options vary in how they respond to this situation:  Both White's flow and the sign flow are ill-defined at such points, and the Toda flow does not have it as a fixed point at all.

In the two-state limit, the dynamics and fixed points of the WWF and the tangent flow are identical to those of a viscous pendulum.  By ``viscous'', we mean that the equations of motion are first-order and inertial effects are negligible.  Like the pendulum, the two-state solutions to both of these flows can be analytically expressed in terms of the Gudermannian function, $\gd {x} = \aasin {(\tanh {x})}$\cite {Robertson97}.

For the WWF, the density of the pendulum's bob, and therefore the speed of its progression along the Gudermannian trajectory, increases like $r^2$.  This faster resolution of higher-energy couplings leads to the renormalization group interpretation of the WWF\cite {GlazekWilson93, GlazekWilson94}.  Note, however, that the arbitrarily slow ``tipping'' dynamics around the unstable fixed points makes this analogy somewhat rough and inexact.  The tangent flow, on the other hand, holds the density of each bob constant, and so the renormalization interpretation is not even approximately applicable.  White's flow\cite {White02} avoids this slow tipping behavior, but introduces its own idiosyncrasies, discussed in section~\ref {WhitesFlow}.

\section {Numerical Implementation Issues} \label {Issues}

\subsection {The Dormand--Prince Method}

A typical first attempt at numerically implementing CUTs is to apply an adaptive Runge--Kutta integrator such as the well-known Dormand--Prince method\cite {DormandPrince80} to equation~\eqref {Flow}.  We will review the Dormand--Prince method only by explaining that this general-purpose fifth-order algorithm iteratively employs a handful of derivative evaluations in order to propagate $\HH$ forward in flow-time by a small step size $h$ while estimating its own error so as to adjust $h'$ for the next step.  Calling it fifth-order means that the calculation error theoretically decreases like $h^5$ as the number of steps increases.

Due to the potentially scarce nature of the matrix of errors $\EE$, we suggest using the entry-wise infinity norm, $\norm {\CCdot}_\infty^\mathrm {entry}$, in order to reduce $\EE$ to a single scalar representing the overall error magnitude:
\begin {equation}
  \norm {\EE}_\infty^\textrm {entry} = \max_{a, b} \left| \EE_{ab} \right|. \label {Norm}
\end {equation}

\subsection {Stiffness} \label {Stiffness}

Upon plugging Wegner's original choice of generator, $\eta = r^2 \sin {2 \theta}$, into the Dormand--Prince method, one finds that the fictitious flow-time $\tau$ must grow surprisingly large before the matrix can be called diagonalized and the flow completed.  This is not an artifact of the implementation; the $r^2$ factor in the generator causes those off-diagonal elements with small radii to decay very slowly.  While one might hope that the integrator would accelerate and begin taking larger steps in this regime, this can not occur because it would lead to over-correction and oscillatory instability in those elements with large radii.  Thus, the integrator is forced to take small steps indefinitely.  This general class of inefficiency is well-known in the numerical analysis literature and is referred to as ``stiffness''.  The most established approach for remediating stiffness is called ``stabilization''.  Stabilized integrators are often implicit, meaning that they require an equation to be numerically solved during the course of each step.\cite {SeinfeldEtAl70, ByrneHindmarsh87}

One can resolve this problem by using a different flow generator, but as we will show in section~\ref {NovelFlow}, none of the previously proposed options fit the bill.  We will therefore construct a generator which resolves this issue.  Furthermore, in sections~\ref {UnitaryIntegrator} and~\ref {Trotter}, we will introduce three integrators which are able to sidestep this issue, even with Wegner's original choice of generator, through two very different mechanisms.

\subsection {Unitarity} \label {Unitarity}

A second flaw which becomes apparent when applying the Dormand--Prince method to CUTs is the slow loss of the unitary similarity relation between $\HH$ and $\HHz$.  This leads to errors in the calculated eigenvalues.  To some extent, such fluctuation is inevitable due to floating-point rounding errors, but the Dormand--Prince method deviates even in the absence of rounding in a step size-dependent manner.

In order to understand this source of error and how to mitigate it, consider the simple system given by $\dot x = -y$ and $\dot y = x$.  These differential equations clearly induce uniform, counterclockwise circular motion about the origin with a constant radius $\sqrt {x^2 + y^2}$.  However, when applied to this system, most numerical integrators will cause the radius to eventually either converge to zero or diverge to infinity in an exponentially spiraling fashion.  Like the family of exponential integrators\cite {HochbruckOstermann10} which were constructed for differential equations with approximately exponential behavior, one could easily write a far superior, specialized integrator for this system, or even a perturbation on it, which uses a rotation matrix in its step propagator in order to avoid this issue.  This broadly applicable technique of exactly solving for the contributions from individual terms in a differential equation separately before combining them is known as ``operator splitting''.\cite {McLachlanQuispel02}

The integrators in sections~\ref{UnitaryIntegrator} and~\ref {Trotter} extend this idea to CUTs and therefore preserve the unitarity similarity, up to rounding error.  In general, integrators which restrict their evolution to some exact submanifold in phase-space are called ``geometric integrators''.  Because the invariant in our case is unitary similarity, our integrators will fall into the subcategory of ``unitary integrators''.

\section {The Uniform Tangent Decay Flow} \label {NovelFlow}

\subsection {White's Flow} \label {WhitesFlow}

We will now provide a possible resolution to the issue of numerical instability and stiffness by constructing an infinitesimal flow generator.  Out of the established flows in table~\ref {Flows}, only White's flow\cite {White02} lacks the stiffening $r$ or $r^2$ factors.  Returning to the two-state limit in section~\ref {TwoState}, \emph {i.e.} neglecting the difficult summed terms in equations~(\ref {FFlow}bc), we can see that this flow generator causes each off-diagonal element to attempt to decay towards zero with the same characteristic time constant.\footnotetext {\label {FNWhite}In fact, White originally proposed using the generator $\eta = j/\delta_0$, using the level spacing from the initial Hamiltonian.\cite {White02}  While sometimes practical, we find this dependence on the starting point of the flow to be conceptually unsatisfying, particularly because $\delta_0$ could well equal zero, and so will not consider it further.}\footnoteref {FNWhite}  At first glance, one might think that the fact that $I_2^J$ decays in exact proportion with $e^{-2 \tau}$ resolves the inefficiency issue, but correctly implementing White's flow with adaptive time steps is also extremely slow.

The severe slowdown in the numerical implementation of White's flow arises at each attempted level crossing, when $\delta$ becomes \emph {very} close to zero, and $\theta$ likewise approaches $\pm \pi/2$.  This causes the generator $\eta = \tan {(\theta)}/2$ to begin to diverge in magnitude, the Hamiltonian to rotate extremely quickly with respect to $\tau$, and the two diagonal elements to repel each other, so as to ultimately prevent $\delta$ from ever actually changing sign.  Only once the Hamiltonian has rotated sufficiently far so as to actually interchange the two relevant axes is this ``avoided crossing'' completed, with $\delta$ retreating back up from zero.  Through this somewhat bizarre and very slow to correctly simulate process, White's flow generally prevents level crossings and maintains the numerical order of the diagonal elements of the Hamiltonian.

\subsection {Derivation of the Tangent Flow}

\begin {figure} [htb]
  {\setlength{\unitlength}{1pt}
\makeatletter%
\let\ASYencoding\f@encoding%
\let\ASYfamily\f@family%
\let\ASYseries\f@series%
\let\ASYshape\f@shape%
\makeatother%
{\catcode`"=12%
\includegraphics{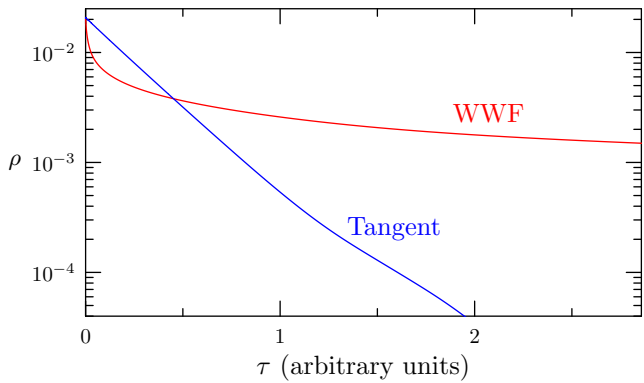}%
}%
\definecolor{ASYcolor}{gray}{0.000000}\color{ASYcolor}
\fontsize{10.037500}{12.045000}\selectfont
\usefont{\ASYencoding}{\ASYfamily}{\ASYseries}{\ASYshape}%
\ASYalignT(-214.238136,42.200112)(-1.000000,-0.500000){0.800000 0.000000 0.000000 0.800000}{\vphantom{$10^4$}$10^{-4}$}%
\definecolor{ASYcolor}{gray}{0.000000}\color{ASYcolor}
\fontsize{10.037500}{12.045000}\selectfont
\ASYalignT(-214.238136,83.824477)(-1.000000,-0.500000){0.800000 0.000000 0.000000 0.800000}{\vphantom{$10^4$}$10^{-3}$}%
\definecolor{ASYcolor}{gray}{0.000000}\color{ASYcolor}
\fontsize{10.037500}{12.045000}\selectfont
\ASYalignT(-214.238136,125.448842)(-1.000000,-0.500000){0.800000 0.000000 0.000000 0.800000}{\vphantom{$10^4$}$10^{-2}$}%
\definecolor{ASYcolor}{gray}{0.000000}\color{ASYcolor}
\fontsize{10.037500}{12.045000}\selectfont
\ASYalignT(-234.882230,83.824477)(-1.000000,-0.188888){1.000000 0.000000 0.000000 1.000000}{\small $\rho$}%
\definecolor{ASYcolor}{gray}{0.000000}\color{ASYcolor}
\fontsize{10.037500}{12.045000}\selectfont
\ASYalignT(-210.474074,21.872049)(-0.500000,-1.000000){0.800000 0.000000 0.000000 0.800000}{\vphantom{$10^4$}$0$}%
\definecolor{ASYcolor}{gray}{0.000000}\color{ASYcolor}
\fontsize{10.037500}{12.045000}\selectfont
\ASYalignT(-136.895976,21.872049)(-0.500000,-1.000000){0.800000 0.000000 0.000000 0.800000}{\vphantom{$10^4$}$1$}%
\definecolor{ASYcolor}{gray}{0.000000}\color{ASYcolor}
\fontsize{10.037500}{12.045000}\selectfont
\ASYalignT(-63.317878,21.872049)(-0.500000,-1.000000){0.800000 0.000000 0.000000 0.800000}{\vphantom{$10^4$}$2$}%
\definecolor{ASYcolor}{gray}{0.000000}\color{ASYcolor}
\fontsize{10.037500}{12.045000}\selectfont
\ASYalignT(-105.362506,11.304875)(-0.500000,-0.750000){1.000000 0.000000 0.000000 1.000000}{$\tau$ (arbitrary units)}%
\begin{picture}( 241.848425, 142.263780)%
\end{picture}%
\kern -241.848425pt%
\definecolor{ASYcolor}{rgb}{1.000000,0.000000,0.000000}\color{ASYcolor}
\fontsize{10.037500}{12.045000}\selectfont
\ASYalign(-58.062300,102.555441)(-0.500000,-0.500000){WWF}%
\definecolor{ASYcolor}{rgb}{0.000000,0.000000,1.000000}\color{ASYcolor}
\fontsize{10.037500}{12.045000}\selectfont
\ASYalign(-93.800233,58.849858)(-0.500000,-0.278481){Tangent}%
}








  \caption {In general, the tangent flow exponentially approaches diagonalization with respect to flow-time, while the WWF slows down.  This particular flow was initialized with a half-filled ten-site MBL Hamiltonian with $W = 1$ as described in section~\ref {MBL}.  The metric $\rho$ is defined in equation~\eqref {Rho}.  Using the Dormand--Prince integrator, calculation time is roughly proportional to flow-time, because the step sizes are bounded, although the relative flow-time scales shown here are arbitrary, reflecting our freedom to arbitrarily rescale the WWF flow-time, which is inversely proportional to the square of the energy scale.}
  \hrule 
  \label {NovelPlot}
\end {figure}

What is needed is some kind of ``softening'' of White's generator when $\theta$ equals $\pm \pi/2$.  Let us attempt to induce the uniform exponential decay not of each $j$, as is the case with White's flow, but instead of each $\tan {\theta} = j/x$.  The standard differential quotient rule tells us that, in the two-state limit,
\begin {equation}
  \dv{\tan {\theta}}{\tau} = \frac {x \dot j - j \dot x} {x^2} = -2 \eta \frac { \left( x^2 + j^2 \right) } {x^2} = -2 \eta \left( \tantan {\theta} + 1 \right).
\end {equation}
This means that our $\eta$ must be proportional to
\begin {equation}
  \frac {2 \tan {\theta}} {\tantan {\theta} + 1} = \sin {2 \theta} = \frac {\delta j} {x^2 + j^2} = \frac {2} {x/j + j/x}.
\end {equation}
Taking this as our choice of generator leads to what we call the ``uniform tangent decay flow'' which conveniently avoids both the stiffness issues exhibited by Wegner's flow and the strange level crossing behavior of White's flow, although level crossings still lead to some amount of reduction in step size.  The tangent flow's exponential approach towards diagonalization is demonstrated for a generic MBL Hamiltonian in figure~\ref {NovelPlot}.

Note that this generator is essentially the same as that of the WWF, but without the $r^2$ factor which quadratically biases those off-diagonal elements with large radii to decay more rapidly.  This prevents the usual renormalization group interpretation discussed in section~\ref {Renormalization} from being applicable to the tangent flow.

\section {Stable Unitary Integrators} \label {UnitaryIntegrator}

\subsection {Introduction}

As mentioned in section~\ref {Unitarity}, the first integrators that we will introduce for numerically implementing CUTs were inspired by the family of exponential integrators\cite {HochbruckOstermann10}, which we will now briefly motivate:  Given an ``almost linear'' differential equation such as $\dot y = \alpha y + \epsilonF {(y)}$ where $\epsilonF {(y)}$ can be regarded as a small perturbation on the first term, one can use our ability to analytically integrate the unperturbed system ($\yF {(\tau)} = \yF {(0)} e^{\alpha \tau}$) in order to take steps which are exactly correct in that limit and then ``add the perturbation back in''.  This frequently allows for highly accelerated performance when compared to the usual linear integrators.  The widely applicable approach of exactly solving for the contributions from multiple parts of the system individually before combining them into a final step operation is known as ``operator splitting''.\cite {McLachlanQuispel02}

Furthermore, our integrators are in the class of geometric integrators.  This means that they exactly preserve a geometric property, in this case the unitary similarity to $\HHz$, up to rounding error.\cite {McLachlanQuispel06, Hairer02}  Symplectic integrators, for instance, take steps that are proper symplectomorphisms, \emph {i.e.} they preserve the symplectic two-form, and are a classic illustration of geometric integrators.\cite {Stuchi02}  While the general idea of developing unitary integrators has been explored before,\cite {DieciEtAl94, ShadwickBuell97, BuonoLopez99, Higham96} our particularly simple approach to stabilizing these integrators specifically to allow them to efficiently handle the stiffness of the WWF, is, to our knowledge, new.

\subsection {First-Order} \label {FirstOrder}

\subsubsection {Integrator Step}

The crux of the idea is to, instead of starting with equation~\eqref {Flow}, go back to equation~\eqref {Exponential}.  Thus, each of the integrators' steps will be governed by the formula
\begin {equation}
  \HHF {(\tau + h)} = e^{h \EEtaF {(\tau)}} \HHF {(\tau)} e^{-h \EEtaF {(\tau)}}. \label {UnitaryStep}
\end {equation}
While the problem of actually computing matrix exponentials is notoriously rife with pitfalls,\cite {MolerLoan03} our situation is particularly painless because $\norm {h \EEta}$ can be assumed to be small.  We therefore need to evaluate only a handful of terms of the infinite series
\begin {equation}
  e^{h \EEta} = \mathlarger{\mathlarger{\sum}}_{k = 0}^\infty \frac {h^k} {k!} \EEta^k \label {ExponentialTaylorSeries}
\end {equation}
before the remaining terms fall below the rounding limit.

In fact, it is not necessary to make use of equation~\eqref {ExponentialTaylorSeries}.  Instead, the order~$(1, 1)$~Pad\'e approximant to the exponential function,
\begin {equation}
  e^{h \EEta} \approx \frac {2 + h \EEta} {2 - h \EEta},
\end {equation}
can be faster to calculate and is perfectly unitary and accurate up to the order of our method.  We have therefore used it in place of the exponential Taylor series.  This approximation is also known as the Cayley transform.\cite {DieleEtAl98}\textsuperscript ,\cite{ShadwickBuell97}  Here, and for the subsequent third-order integrator, the choice of matrix exponential algorithm can be critical to performance, and experimentation is advised, taking the specific demands and available implementations into account.

\subsubsection {Stabilization} \label {Correction}

This algorithm already preserves unitarity, but it is known in the literature, and is still susceptible to the instability that results from the stiffness of the WWF.  To resolve this issue, we will take advantage of our ability to exactly integrate the flow differential equations in the two-state limit from section~\ref {TwoState}.  First, we analytically solve equation~\eqref {RotationRate}, namely $\dot \theta = -2 \eta$, with $\thetaF {(\tau)}$ properly initialized as in our flow.  Then we evaluate $\thetaF {(\tau + h)}$ and use the result to calculate an effective value, constant for the duration of the step, to use as a corrected generator according to
\begin {equation}
  \eta_h = \frac {\thetaF {(\tau)} - \thetaF {(\tau + h)}} {2h}.
\end {equation}
This correction is applied to each independent component of the generator separately, as if it were the only active rotation.

In the case of the WWF, this gives
\begin {samepage}\begin {align} \begin {split}
  \eta_h =& \, \frac {\theta - \aatan { \left( \frac {\tan {\theta}} {e^{4 r^2 h}} \right) }} {2h} \\
  =& \, \frac {\aatan { \left( \frac {j} {x} \right) } - \aatan { \left( \frac {j} {x e^{4 \left( x^2 + j^2 \right) h}} \right) }} {2h}.
\end {split} \end {align}\end {samepage}
The large-radius off-diagonal element instability is prevented because this corrected generator blocks the flow from mistakenly rotating ``too far'' and changing the sign of $j$, in the two-state limit.  This allows the step sizes to increase as the flow progresses towards diagonalization and slows down due to the small-radius off-diagonal elements.  On the other hand, the unstable Dormand--Prince integrator's step size is bounded from above by the large-radius elements.

For our uniform tangent decay flow, we obtain
\begin {equation}
  \eta_h = \frac {\theta - \aatan { \left( \frac {\tan {\theta}} {e^{4h}} \right) }} {2h}  = \frac {\aatan { \left( \frac {j} {x} \right) } - \aatan { \left( \frac {j} {x e^{4h}} \right) }} {2h}.
\end {equation}
This correction is not truly necessary because the flow itself is already not stiff, but it can still improve performance.

\subsubsection {Adaptive Step Sizes}

In order to develop this into an adaptive integrator, it is necessary for it to estimate its own error at each step in order to adjust its next step size.  Because we can assume that each step is a true unitary rotation, all of the nonrounding error can be attributed to the change of $\EEta$ during that step and is on the order of $h^2 \dot \EEta$.  The error is caused by our failure to predict how $\EEta$ changes over the course of the step, and we wish to limit this relative to the ``rotation rate'', $\norm {\EEta}_F$.  Therefore, we can calculate the subsequent step size according to
\begin {equation}
  h' = \frac {\epsilon h} {n} \frac {\norm {\EEtahF {(\tau)}}_F} {\norm {\EEtazF {(\tau + h)} - \EEtahpF {(\tau)}}_\infty^\mathrm {entry}}, \label {StepSizeUpdate}
\end {equation}
where $\epsilon$ is an adjustable tolerance parameter, $\EEtaz$ is the uncorrected flow generator, and $\EEtahp$ is our approximation's prediction for the value of $\EEta$ at the end of the step.  For the WWF this is given by
\begin {equation}
  \eta_h' = \frac {\delta j r^2 e^{4 r^2 h}} {x^2 + j^2 e^{8 r^2 h}},
\end {equation}
and the prediction for the tangent flow is the same, after dropping the three $r^2$ factors.  However, due to its uniform exponential decay, one may find constant step sizes to be a superior choice when implementing the tangent flow. 

With most adaptive integrators, it is standard to limit the ratio of each pair of consecutive step sizes so as to prevent them from changing too rapidly.  For example, one can require that that their ratio lie between one half and two.  Furthermore, if $h'$ is less than some fixed fraction of $h$, such as $3h/4$, one should usually repeat the step with the new, more conservative value.

\subsection {Third-Order} \label {ThirdOrder}

\subsubsection {Introduction}

While the above integrator performs fairly well, it is only first-order and so can not compete with the higher-order Runge--Kutta algorithms' efficiency during the early stages of the flow.  Higher-order algorithms provide better-than-linear returns on precision when decreasing the step size, and so it was desirable for us to devise a third-order extension which retains the nice stability and unitarity properties.  Our approach falls into the class of multiderivative methods, which means we calculate not only $\EEta$ at each step, but also its first two derivatives.  Using these to construct an effective, stabilized generator requires taking multiple types of corrections into account, as explained below.

The techniques in this section are applicable to most well-behaved choices of generator.  For simplicity's sake, however, we will focus on the classic WWF and the uniform tangent decay flow from section~\ref {NovelFlow}.

\subsubsection {Integrator Step}

We begin with the basic WWF equations $\EEta = \comm {\HH_\mathrm {Diag.}} {\HH}$ and $\dot \HH = \comm {\EEta} {\HH}$.  These immediately allow us to calculate that
\begin {samepage}\begin {subequations}\label {EtaDots}\begin {align}
  \dot \EEta = & \, \comm {\dot \HH_\mathrm {Diag.}} {\HH} + \comm {\HH_\mathrm {Diag.}} {\dot \HH},& \label {EtaDot} \\
  \ddot \HH = & \, \comm {\dot \EEta} {\HH \vphantom {\dot \HH}} + \comm {\EEta} {\dot \HH}, \\
\intertext {and}
  \ddot \EEta = & \, \comm {\ddot \HH_\mathrm {Diag.}} {\HH} + 2 \comm {\dot \HH_\mathrm {Diag.}} {\dot \HH} + \comm {\HH_\mathrm {Diag.}} {\ddot \HH}. \label {EtaDoubleDot}
\end {align}\end {subequations}\end {samepage}
In terms of the elements, $\eta = \delta j$, $\dot \eta = \dot \delta j + \delta \dot j$, and $\ddot \eta = \ddot \delta j + 2 \dot \delta \dot j + \delta \ddot j$.  In the two-state limit from section~\ref {TwoState}, we could easily calculate a higher-order unitary step operator according to $e^{h \ZZetaF {(h)}}$, where
\begin {equation}
  \ZZetaF {(h)} = \frac 1 h \int_0^h \EEta + \dot \EEta \tau + \frac {\ddot \EEta} 2 \tau^2 \dt = \EEta + \frac {\dot \EEta} 2 h + \frac {\ddot \EEta} 6 h^2
\end {equation}
is an effective generator corrected by the derivatives.

However, for larger systems, the noncommutativity of $\EEta$ and its derivatives must also be taken into account according to the Magnus expansion:\cite {ShadwickBuell97, ShadwickBuell01}  While $\ZZetaF {(0)} = \ZZetaz = \EEta$, and the derivative with respect to the step size $h$, $\ZZetazp = \ZZetapF {(0)} = \dot \EEta/2$,
\begin {equation}
  \ZZetazpp = \frac {2 \ddot \EEta - \comm {\EEta} {\dot \EEta}} 6. \label {Magnus}
\end {equation}
$\ZZetaF {(h)}$ can now be correctly expressed as a second-order Maclaurin series using these corrected coefficients:
\begin {equation}
  \zetaF {(h)} = \zeta_0 + \zeta_0' h + \frac {\zeta_0''} 2 h^2 \label {UncorrectedZeta}
\end {equation}

In order to finish upgrading to a third-order algorithm, the degree of the Pad\'e approximant must also be increased to $(2, 2)$:
\begin {equation}
  e^{h \ZZeta} \approx \frac {12 + 6h \ZZeta + h^2 \ZZeta^2} {12 - 6h \ZZeta + h^2 \ZZeta^2}.
\end {equation}

\subsubsection {Stabilization} \label {ThirdOrderStabilization}

This algorithm is already third-order, but without stabilization, the multiderivative polynomial extrapolation makes it highly unstable.  We suggest stabilizing it through the observation that $\eta$ decays asymptotically like $e^{-4 r^2 \tau}$.  Thus, we make the ansatz that
\begin {equation}
  h \zetaroF {(h)} = \bigintssss_0^h \frac {\left( c_0 + c_1 \tau + \frac {c_2} 2 \tau^2 \right)} {e^{4 r_0^2 \tau}} \dt, \label {Integral}
\end {equation}
where $r_0$ is the radius at the beginning of the step.  Evaluating this integral analytically gives
\begin {equation} \begin {gathered}
  h \zetaroF {(h)} = \frac 1 {64 r_0^6} \bigg[ 16 c_0 r_0^4 + 4 c_1 r_0^2 + c_2 \\
  - \frac {16 c_0 r_0^4 + 4 c_1 r_0^2 + c_2 + \left( 16 c_1 r_0^4 + 4 c_2 r_0^2 \right) h + 8 c_2 r_0^4 h^2} {e^{4 r_0^2 h}} \bigg]. \label {Evaluated}
\end {gathered} \end {equation}
Matching terms in the corresponding Maclaurin series,
\begin {equation}
  \zetaroF {(h)} \approx c_0 + \frac {-4 c_0 r_0^2 + c_1} 2 h + \frac {16 c_0 r_0^4 - 8 c_1 r_0^2 + c_2} 6 h^2,
\end {equation}
with equation~\eqref {UncorrectedZeta} allows us to find the three $c$ coefficients to be
\begin {samepage}\begin {subequations}\begin {alignat} {6}
  c_0 = &&& \zeta_0, \\
  c_1 = && \, 4 r_0^2 & \zeta_0 && + & 2 & \zeta_0', \\
\intertext {and}
  c_2 = && \, 16 r_0^4 & \zeta_0 && + & \, 16 r_0^2 & \zeta_0' && + & \, 3 & \zeta_0''.
\end {alignat}\end {subequations}\end {samepage}
Plugging the above results into equation~\eqref {Evaluated} ultimately results in an expression for the stabilized $\zeta_{r_0}$ in terms of $\EEta$ and its derivatives:
\begin {equation} \begin {gathered}
  h \zetaroF {(h)} = \frac 1 {128 r_0^6} \bigg\{ 96 r_0^4 \eta + 24 r_0^2 \dot \eta + 2 \ddot \eta - \comm {\EEta} {\dot \EEta} \\
  - e^{-4 r_0^2 h} \Big[ 96 r_0^4 \eta + 24 r_0^2 \dot \eta + 2 \ddot \eta - \comm {\EEta} {\dot \EEta} \\
  + \big( 256 r_0^6 \eta + 96 r_0^4 \dot \eta + 8 r_0^2 \ddot \eta - 4 r_0^2 \comm {\EEta} {\dot \EEta} \big) h \\
  + \big( 256 r_0^8 \eta + 128 r_0^6 \dot \eta + 16 r_0^4 \ddot \eta - 8 r_0^4 \comm {\EEta} {\dot \EEta} \big) h^2 \Big] \bigg\}.
\end {gathered} \end {equation}
When the exponent $4 r_0^2 h$ is small, care must be taken to avoid catastrophic cancellation during the evaluation of these corrected expressions.  In this limit, an uncorrected Taylor series such as equation~\eqref {UncorrectedZeta} can be used instead.

\subsubsection {Adaptive Step Sizes}

The error of each step is now on the order of $h^4 \dddot \EEta$.  We suggest simply adding a cube root to equation~\eqref {StepSizeUpdate} and adjusting the step sizes according to
\begin {equation}
  h' = h \left( \frac {\epsilon} {n} \frac {\norm {\ZZetaF {(\tau)}}_F} {\norm {\EEtazF {(\tau + h)} - \EEtahpF {(\tau)}}_\infty^\mathrm {entry}} \right)^{1/3},
\end {equation}
where the predicted generator
\begin {equation}
  \eta_h' = \frac {2\eta + \left( 8 \eta r_0^2 + 2 \dot \eta \right) h + \left( 16 \eta r_0^4 + 8 \dot \eta r_0^2 + \ddot \eta \right) h^2} {2 e^{4 r_0^2 h}} \label {ThirdOrderCorrection}
\end {equation}
is the integrand of equation~\eqref {Integral} when $\tau = h$ without the Magnus correction commutator term in equation~\eqref {Magnus}.\footnotetext {\label {FNMagnus}Note that we are implicitly assuming that the Magnus correction does not affect the validity of the decay ansatz in equation~\eqref {Integral}.}\footnoteref {FNMagnus}

\subsubsection {Alternative Flows}

This third-order integrator can be easily extended to alternative choices of generator by modifying equations~\eqref {Flow}, (\ref {EtaDots}ac), and the decay ansatz in equation~\eqref {Integral}.  However, the efficiency of the multiderivative approach may be negatively impacted.

For example, to implement the uniform tangent decay flow from section~\ref {NovelFlow}, we calculate that $\eta = \sin 2 \theta$, $\dot \eta = 2 \dot \theta \cos {2 \theta}$, and $\ddot \eta = 2 \ddot \theta \cos {(2 \theta)} - 4 \dot \theta^2 \sin {(2 \theta)}$.  These can be expressed in terms of the elements of the Hamiltonian matrix using $\theta = \aatan {j/x}$, but the resulting expressions are somewhat lengthy and will be omitted.  Additionally, one must modify the decay ansatz by setting $r$ and $r_0$ to unity in equations~\eqref {Integral} through~\eqref {ThirdOrderCorrection}. 

Note that simulating the tangent flow to third-order precision is a computationally expensive process for many physically inspired initial Hamiltonians.  In particular, small-radius level crossings, where $x$ changes sign while $j \approx 0$, involve rapid fluctuations in $\eta$ which can require very small step sizes to correctly integrate. 

\subsection {Results}

\begin {figure} [htb]
  {\setlength{\unitlength}{1pt}
\makeatletter%
\let\ASYencoding\f@encoding%
\let\ASYfamily\f@family%
\let\ASYseries\f@series%
\let\ASYshape\f@shape%
\makeatother%
{\catcode`"=12%
\includegraphics{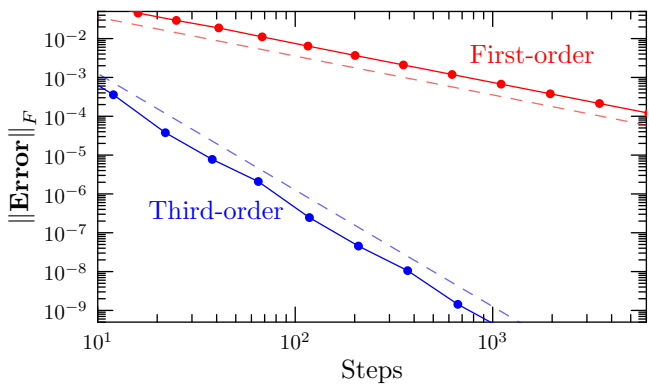}%
}%
\definecolor{ASYcolor}{gray}{0.000000}\color{ASYcolor}
\fontsize{10.037500}{12.045000}\selectfont
\usefont{\ASYencoding}{\ASYfamily}{\ASYseries}{\ASYshape}%
\ASYalignT(-211.538366,28.838964)(-1.000000,-0.500000){0.800000 0.000000 0.000000 0.800000}{\vphantom{$10^4$}$10^{-9}$}%
\definecolor{ASYcolor}{gray}{0.000000}\color{ASYcolor}
\fontsize{10.037500}{12.045000}\selectfont
\ASYalignT(-211.538366,43.538837)(-1.000000,-0.500000){0.800000 0.000000 0.000000 0.800000}{\vphantom{$10^4$}$10^{-8}$}%
\definecolor{ASYcolor}{gray}{0.000000}\color{ASYcolor}
\fontsize{10.037500}{12.045000}\selectfont
\ASYalignT(-211.538366,58.238709)(-1.000000,-0.500000){0.800000 0.000000 0.000000 0.800000}{\vphantom{$10^4$}$10^{-7}$}%
\definecolor{ASYcolor}{gray}{0.000000}\color{ASYcolor}
\fontsize{10.037500}{12.045000}\selectfont
\ASYalignT(-211.538366,72.938582)(-1.000000,-0.500000){0.800000 0.000000 0.000000 0.800000}{\vphantom{$10^4$}$10^{-6}$}%
\definecolor{ASYcolor}{gray}{0.000000}\color{ASYcolor}
\fontsize{10.037500}{12.045000}\selectfont
\ASYalignT(-211.538366,87.638454)(-1.000000,-0.500000){0.800000 0.000000 0.000000 0.800000}{\vphantom{$10^4$}$10^{-5}$}%
\definecolor{ASYcolor}{gray}{0.000000}\color{ASYcolor}
\fontsize{10.037500}{12.045000}\selectfont
\ASYalignT(-211.538366,102.338327)(-1.000000,-0.500000){0.800000 0.000000 0.000000 0.800000}{\vphantom{$10^4$}$10^{-4}$}%
\definecolor{ASYcolor}{gray}{0.000000}\color{ASYcolor}
\fontsize{10.037500}{12.045000}\selectfont
\ASYalignT(-211.538366,117.038199)(-1.000000,-0.500000){0.800000 0.000000 0.000000 0.800000}{\vphantom{$10^4$}$10^{-3}$}%
\definecolor{ASYcolor}{gray}{0.000000}\color{ASYcolor}
\fontsize{10.037500}{12.045000}\selectfont
\ASYalignT(-211.538366,131.738072)(-1.000000,-0.500000){0.800000 0.000000 0.000000 0.800000}{\vphantom{$10^4$}$10^{-2}$}%
\definecolor{ASYcolor}{gray}{0.000000}\color{ASYcolor}
\fontsize{10.037500}{12.045000}\selectfont
\ASYalignT(-232.182460,83.213352)(-0.500000,0.179400){0.000000 1.000000 -1.000000 0.000000}{$\norm{\mathbf {Error}}_F$}%
\definecolor{ASYcolor}{gray}{0.000000}\color{ASYcolor}
\fontsize{10.037500}{12.045000}\selectfont
\ASYalignT(-207.774304,20.649799)(-0.500000,-1.000000){0.800000 0.000000 0.000000 0.800000}{\vphantom{$10^4$}$10^{1}$}%
\definecolor{ASYcolor}{gray}{0.000000}\color{ASYcolor}
\fontsize{10.037500}{12.045000}\selectfont
\ASYalignT(-133.075935,20.649799)(-0.500000,-1.000000){0.800000 0.000000 0.000000 0.800000}{\vphantom{$10^4$}$10^{2}$}%
\definecolor{ASYcolor}{gray}{0.000000}\color{ASYcolor}
\fontsize{10.037500}{12.045000}\selectfont
\ASYalignT(-58.377566,20.649799)(-0.500000,-1.000000){0.800000 0.000000 0.000000 0.800000}{\vphantom{$10^4$}$10^{3}$}%
\definecolor{ASYcolor}{gray}{0.000000}\color{ASYcolor}
\fontsize{10.037500}{12.045000}\selectfont
\ASYalign(-104.012621,10.082625)(-0.500000,-0.778481){Steps}%
\begin{picture}( 241.848425, 142.263780)%
\end{picture}%
\kern -241.848425pt%
\definecolor{ASYcolor}{rgb}{1.000000,0.000000,0.000000}\color{ASYcolor}
\fontsize{10.037500}{12.045000}\selectfont
\ASYalign(-43.437893,125.858123)(-0.500000,-0.500000){First-order}%
\definecolor{ASYcolor}{rgb}{0.000000,0.000000,1.000000}\color{ASYcolor}
\fontsize{10.037500}{12.045000}\selectfont
\ASYalign(-162.955283,67.058633)(-0.500000,-0.500000){Third-order}%
}

    








  \caption {The third-order integrator requires far fewer steps to converge to the true WWF than the first-order one.  The dashed lines show the slope theoretically predicted by their order.  These flows were initialized for a half-filled ten-site MBL Hamiltonian with $W = 1$ as explained in section~\ref {MBL} and run until $\tau = 1$.  The error matrix was estimated by subtracting the resulting $\HHF {(1)}$ from the output of a high-precision flow calculation.}
  \hrule
  \label {OrderPlot}
\end {figure}
As shown in figure~\ref {OrderPlot}, these implementations of the WWF are in fact accurate to their predicted theoretical order.  Where applicable, we recommend that the third-order integrator be used for the numerical implementation of CUTs.  While we have not judged it as worthwhile, one could consider devising even higher-order extensions.

\section {Quantized Trotter Integrator} \label {Trotter}

\subsection {Introduction}

Next, we will develop another stable, unitary integrator.  While all of the integrators introduced in this paper involve operator splitting\cite {McLachlanQuispel02}, this one takes it much further than the previous.  We seek an integrator which puts ``just the right amount'' of computational effort into each pair of off-diagonal elements, in proportion to the rate of rotation between them as set by the generator of the flow.

One lesser-known class of integrators which function along these lines is known as the quantized state system simulators.  Instead of basing the construction of the algorithm on the slicing of time into discrete steps as is typically done, these methods quantize the configuration-space and then calculate when the system would be expected to switch from one state ``cell'' to another.\cite {CellierEtAl08, MigoniEtAl11}  We derived our specialized integrator with this idea and the Trotter decomposition\cite {Trotter59} in mind, while seeking to retain the unitary exactness of the integrators in the previous section.  However, due to the noncommutativity of the rotation generators, our partitioning of configuration-space into cells is not invariant over the course of the flow.

\subsection {Integrator Step}

We begin by decomposing $\EEta$ into a real linear combination of the $(n^2 - n)/2$ generators of the $\oF {(n)}$ Lie algebra.  We will denote these generators as $\OO_{ab} = \delta_{ac}\delta_{bd} - \delta_{ad}\delta_{bc}$ for all $1 \le a < b \le n$.  Note that $e^{\theta \OO_{ab}}$ represents a rotation of $\theta$ radians between the $\ket a$- and $\ket b$-axes, holding all others fixed.

A trivial decomposition is given by
\begin {equation}
  \EEta = \sum_{a < b} \EEta_{ab} \OO_{ab}.
\end {equation}
Referring back to equation~\eqref {UnitaryStep}, the unitary operator which we seek to implement is $e^{h \EEta}$.  The specific form of the Trotter decomposition which we will base our construction on is the first three factors of the Zassenhaus formula:\cite {Magnus54}
\begin {equation}
  e^{\epsilon (\AAA + \BB)} = e^{\epsilon \AAA} e^{\epsilon \BB} e^{\frac {\epsilon^2} 2 \comm {\BB} {\AAA}} \ldots . \label {Zassenhaus}
\end {equation}
From this we can immediately write that, for small $h$,
\begin {equation}
  e^{h \EEta} \approx \prod_{a < b} e^{h \EEta_{ab} \OO_{ab}},
\end {equation}
but we can do better.  Instead of performing each rotation in the proper proportion at every step, we will instead fix the magnitude of each rotation to be some small angle $\iota$, and perform the rotations in an interleaved manner where each occurs with the proper ``frequency'' with respect to $\tau$.  The choice of a fixed angle is convenient not only because it makes the numerical implementation more efficient, but also because it bounds the magnitude of the dominant error introduced by the Trotter decomposition as given by the final $e^{\epsilon^2 \comm {\BB} {\AAA}/2}$ factor in equation~\eqref {Zassenhaus}.

\begin {figure} [htb]
  {\setlength{\unitlength}{1pt}
\makeatletter%
\let\ASYencoding\f@encoding%
\let\ASYfamily\f@family%
\let\ASYseries\f@series%
\let\ASYshape\f@shape%
\makeatother%
{\catcode`"=12%
\includegraphics{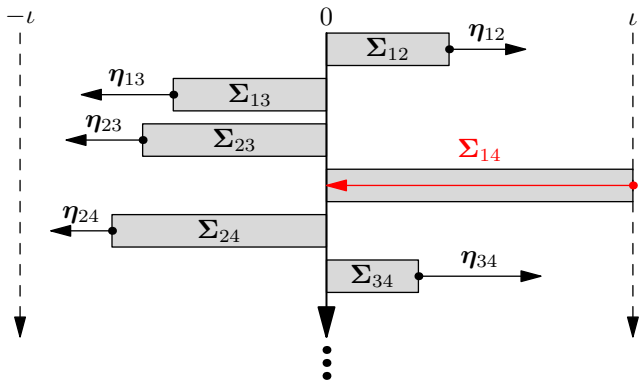}%
}%
\definecolor{ASYcolor}{gray}{0.000000}\color{ASYcolor}
\fontsize{10.037500}{12.045000}\selectfont
\usefont{\ASYencoding}{\ASYfamily}{\ASYseries}{\ASYshape}%
\ASYalign(-118.979763,134.292245)(-0.500000,0.000000){$0$}%
\definecolor{ASYcolor}{gray}{0.000000}\color{ASYcolor}
\fontsize{10.037500}{12.045000}\selectfont
\ASYalign(-3.074560,134.292245)(-0.500000,0.000000){$\iota$}%
\definecolor{ASYcolor}{gray}{0.000000}\color{ASYcolor}
\fontsize{10.037500}{12.045000}\selectfont
\ASYalign(-234.884965,134.292245)(-0.500000,0.125000){$-\iota$}%
\definecolor{ASYcolor}{gray}{0.000000}\color{ASYcolor}
\fontsize{10.037500}{12.045000}\selectfont
\ASYalign(-95.798722,125.145043)(-0.500000,-0.320600){$\mathbf \Sigma_{12}$}%
\definecolor{ASYcolor}{gray}{0.000000}\color{ASYcolor}
\fontsize{10.037500}{12.045000}\selectfont
\ASYalign(-58.129531,128.156293)(-0.500000,0.311111){$\bm \eta_{12}$}%
\definecolor{ASYcolor}{gray}{0.000000}\color{ASYcolor}
\fontsize{10.037500}{12.045000}\selectfont
\ASYalign(-147.956063,107.964380)(-0.500000,-0.320600){$\mathbf \Sigma_{13}$}%
\definecolor{ASYcolor}{gray}{0.000000}\color{ASYcolor}
\fontsize{10.037500}{12.045000}\selectfont
\ASYalign(-194.318144,110.975630)(-0.500000,0.311111){$\bm \eta_{13}$}%
\definecolor{ASYcolor}{gray}{0.000000}\color{ASYcolor}
\fontsize{10.037500}{12.045000}\selectfont
\ASYalign(-153.751323,90.783716)(-0.500000,-0.320600){$\mathbf \Sigma_{23}$}%
\definecolor{ASYcolor}{gray}{0.000000}\color{ASYcolor}
\fontsize{10.037500}{12.045000}\selectfont
\ASYalign(-203.011034,93.794966)(-0.500000,0.311111){$\bm \eta_{23}$}%
\definecolor{ASYcolor}{rgb}{1.000000,0.000000,0.000000}\color{ASYcolor}
\fontsize{10.037500}{12.045000}\selectfont
\ASYalign(-61.027161,82.750254)(-0.500000,0.179400){$\mathbf \Sigma_{14}$}%
\definecolor{ASYcolor}{gray}{0.000000}\color{ASYcolor}
\fontsize{10.037500}{12.045000}\selectfont
\ASYalign(-159.546584,56.422389)(-0.500000,-0.320600){$\mathbf \Sigma_{24}$}%
\definecolor{ASYcolor}{gray}{0.000000}\color{ASYcolor}
\fontsize{10.037500}{12.045000}\selectfont
\ASYalign(-211.703925,59.433639)(-0.500000,0.311111){$\bm \eta_{24}$}%
\definecolor{ASYcolor}{gray}{0.000000}\color{ASYcolor}
\fontsize{10.037500}{12.045000}\selectfont
\ASYalign(-101.593982,39.241725)(-0.500000,-0.320600){$\mathbf \Sigma_{34}$}%
\definecolor{ASYcolor}{gray}{0.000000}\color{ASYcolor}
\fontsize{10.037500}{12.045000}\selectfont
\ASYalign(-61.027161,42.252975)(-0.500000,0.311111){$\bm \eta_{34}$}%
}

  \caption {In the Trotter integrator, each pair of off-diagonal elements in the accumulator $\SSigma$ grows at the rate indicated by the corresponding element of $\EEta$.  One can calculate a ``hitting time'' for each of them when that $\sigmaF {(\tau)}$ will reach $\pm \iota$.  To perform each step, we find which element will hit next, jump forward to that flow-time, perform a Jacobi rotation of magnitude $\mp \iota$ between the appropriate basis axes, and reset that $\sigma$ to zero.  Here, a Jacobi rotation of magnitude $-\iota$ between the $\ket 1$- and $\ket 4$-axes is indicated in red, which will then require recalculating all of the shown generators and expected hitting times except for the $\OO_{23}$ element.}
  \hrule
  \label {SchematicPlot}
\end {figure}
More formally, we can imagine integrating $\EEta$ into an accumulator $\SSigma$, which is initially the zero matrix.  Again using the entry-wise infinity norm defined in equation~\eqref {Norm}, when $\norm {\SSigma}_\infty^\mathrm {entry}$ grows to reach $\iota$, we can generally say that this was caused by a single pair of elements $\SSigma_{ab} = -\SSigma_{ba} = \pm \iota$.  This triggers a Jacobi rotation\cite {Kelley95} of magnitude $\mp \iota$ between the $\ket a$- and $\ket b$-axes.  Next, $\SSigma_{ab}$ and $\SSigma_{ba}$ are reset to zero and the integration continues until the entry-wise infinity norm again reaches $\iota$.  See figure~\ref {SchematicPlot} for a schematic illustration.

Note that the step sizes are generally distributed in a Poissonian fashion at any given point in the flow.  The instantaneous step rate is roughly proportional to the entry-wise one norm of the generator:
\begin {equation}
  \frac {\norm {\EEta}_1^\mathrm {entry}} \iota = \frac 1 \iota \mathlarger{\mathlarger{\sum}}_{a < b} \left| \EEta_{ab} \right|.
\end {equation}

\subsection {Stabilization}

At the end of the flow, some care must be taken to ensure that $\iota$ does not cause us to rotate ``too far'':  $j$ can be set to zero as in the usual Jacobi eigenvalue algorithm, but it should not be caused to change sign.

Alternatively, if we are willing to give up the efficiency boost allowed by the use of a fixed $\iota$, we can ignore this consideration and instead always rotate by a corrected $\iota_\theta$ value calculated in a manner similar to the corrected generator $\eta_h$ in section~\ref {Correction}.  For both the WWF and the uniform tangent decay flow from section~\ref {NovelFlow}, analytically solving the two-state differential equation gives us a corrected Jacobi rotation angle of
\begin {samepage}\begin {align} \begin {split}
    \iota_\theta =& \, \frac {\aatan { \left( \tan {\theta} \, e^{4 \iota \csc {2 \theta}} \right) } - \theta} {2} \\
    =& \, \frac {\aatan { \left( \frac {j} {x} e^{ 2 \iota \frac { x^2 + j^2 } {xj}} \right) } - \aatan { \left( \frac {j} {x} \right) }} {2}.
\end {split} \end {align}\end {samepage}
This strategy appears to reduce performance slightly.

\subsection {Implementation}

The key feature of this algorithm is that the computational expense is approximately proportional to the rotational ``path length'' of the basis during the course of the flow.  This is why it is immune to the features of the WWF which cause certain other integrators to perform poorly.

This approach can be conveniently implemented by noting that $\EEta$ is now piecewise constant, assuming it does not depend explicitly on $\tau$, and so $\SSigma$ is piecewise linear.  Each pair of off-diagonal elements in $\SSigma$ has an expected ``hitting time'' of $\tau + (\sgn {(\eta)} \iota - \sigma)/\eta$ to reach $\pm \iota$.  We can efficiently keep track of these in a bimap-like tree-based data structure, and then read off the next indicated rotation from the first leaf of the tree.  After each rotation, only $2n - 3$ generator values and hitting times need to be updated.

If one desires to halt the flow at a particular time $\tau$, it is necessary to iterate until just before that value is surpassed and then ``clear out'' the queue from the beginning to the end, rotating by only the appropriate fraction of $\iota$ each time.

In terms of raw performance, we found this algorithm to be inferior to the third-order one in section~\ref {ThirdOrder} but still far superior to the Dormand-Prince method.  Regardless, we feel that it is fundamentally different enough that it may prove useful in future work.

It seems likely that this algorithm could be parallelized in essentially the same manner as the Jacobi eigenvalue algorithm\cite {MargarisEtAl14}.  Additionally, we suspect that higher-order extensions are possible, such as treating $\EEta$ as piecewise linear instead of piecewise constant.

\section {Application to Many-Body Localization} \label {MBL}

\subsection {The MBL Model}

We tested the above CUT integrators by applying them to a standard model of one-dimensional MBL on a periodic lattice\cite {RademakerOrtuno16}.  The Hamiltonian can be written in terms of second-quantized spinless fermions as
\begin {equation}
  \HH = \mathlarger{\sum}_{k = 1}^L \left[ \mu_k n_k + V n_k n_{k + 1} + \left( t c_k^\dagger c_{k + 1} + \textrm {h.c.} \right) \right],
\end {equation}
where $n_k = c_k^\dagger c_k$, and each $\mu_k$ was drawn randomly from the uniform distribution over the range $[-W/2, W/2]$.  The three terms correspond to on-site potentials with quenched disorder, interactions of strength $V = 1$, and hopping at the rate $t = 1$, respectively.

As $W$ is increased, this system is believed to undergo a transition from an ergodic extended phase to a nonergodic localized one, possibly passing through an intermediate nonergodic extended phase.\cite {SerbynMoore16}  Our understanding of these transitions is hampered by finite-size effects controlled by $L$.  This is particularly problematic for the computational study of MBL because the dimensionality of the Hilbert space depends exponentially on $L$.

We set $L$ to be ten and, as the Hamiltonian preserves the total particle number, restrict our attention to the half-filled sector of the Hilbert space containing five particles.  Therefore $n$, the dimension of our matrices, is $\binom {10} {5} = 252$.

Before beginning the flow, we are free to pre-rotate our Hamiltonian into whatever basis we prefer.  We found that the easily calculable single-particle localized basis which diagonalizes the free Hamiltonian with $V = 0$ gives a significant performance boost.  Not only does this pre-rotation immediately reduce the $\rho$ metric from equation~\eqref {Rho} by almost an order of magnitude, it also smooths and accelerates the subsequent flow.

\subsection {Level Repulsion Metric}

In order to visualize the concept of Hilbert space percolation present in the MBL literature\cite {PotterEtAl15} using CUTs, we constructed a metric that measures how much ``work'' the decay of any given pair of off-diagonal elements performs towards inducing the repulsion of the values of the diagonal elements which they connect.  Formally, we integrated each $\dot {x^2}$ in the two-state limit from section~\ref {TwoState}:
\begin {samepage}\begin {equation} \begin {aligned}
  \XXi_{ab} = & \bigintssss_0^\infty \left. \dv{\XX_{ab}^2}{\tau} \right|_{\EEta = \eta \OO_{ab}} \dt \\
  = & \int_0^\infty \left. 2 \XX_{ab} \dot \XX_{ab} \right|_{\EEta = \eta \OO_{ab}} \dt.
\end {aligned} \end {equation}\end {samepage}
If we were not in the two-state limit, this would simply equal $x_\infty^2 - x_0^2$, where $x_\infty$ is taken from the diagonalized matrix.  Instead, equation~\eqref {XFlow} becomes just $\dot x = 2 \eta j$, so
\begin {equation}
  \Xi = \int_0^\infty 4 \eta x j \dt.
\end {equation}
For the WWF, which we will use in this section, the final integrand equals $2 \eta^2$.

In some sense, $\XXi_{ab}$ measures the total amount of repulsion which has occurred between $D_a$ and $D_b$ due to the decay of $\JJ_{ab}$ throughout the course of the flow.  One can expect that $\XXi$ will be approximately sparse if and only if the system is localized.  This is because localized Hamiltonians do not effectively mix most pairs of states.  This has the consequence that the eigenvalues in the extended phase obey random matrix, Wigner--Dyson, Gaussian orthogonal ensemble statistics\cite {Mehta04}, but behave in a Poissonian manner in the localized phase.  The intermediate phase is believed to obey some sort of power-law repulsion eigenvalue statistics.\cite {SerbynMoore16}

Furthermore, assuming the flow succeeds in diagonalizing the Hamiltonian, one can easily show that
\begin {equation}
  \sum_{a \ne b} {\XXi_{ab}} = \sum_{a \ne b} {\JJFab {(\tau = 0)}^2} = I_2^J {(\tau = 0)},
\end {equation}
the initial value of the metric defined in equation~\eqref {I2J}.  The various phases of localization phenomena correspond analogously to qualitatively distinct economic systems for allocating the scarce initial $I_2^J$ resource amongst the $\Xi$ elements.

It is generally not difficult to extend an integrator to keep track of the matrix $\XXi$.  For example, the third-order WWF integrator from section~\ref {ThirdOrder} can be extended by symbolically calculating the $\Delta \Xi$ increment integral and then numerically evaluating the resulting expression at every step for each pair of off-diagonal elements using the same decay ansatz and numerical techniques as in section~\ref {ThirdOrderStabilization}.

\subsection {Results}
\begin {figure} [htb]
  {\setlength{\unitlength}{1pt}
\makeatletter%
\let\ASYencoding\f@encoding%
\let\ASYfamily\f@family%
\let\ASYseries\f@series%
\let\ASYshape\f@shape%
\makeatother%
{\catcode`"=12%
\includegraphics{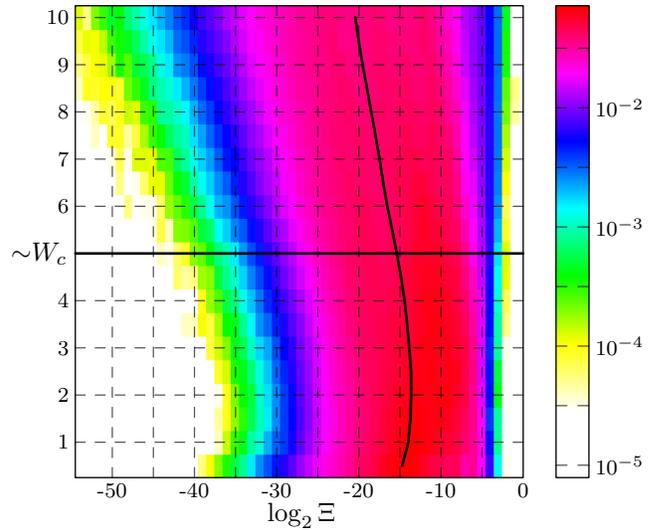}%
}%
\definecolor{ASYcolor}{gray}{0.000000}\color{ASYcolor}
\fontsize{8.030000}{9.636000}\selectfont
\usefont{\ASYencoding}{\ASYfamily}{\ASYseries}{\ASYshape}%
\ASYalign(-218.645310,33.686765)(-1.000000,-0.500000){1}%
\definecolor{ASYcolor}{gray}{0.000000}\color{ASYcolor}
\fontsize{8.030000}{9.636000}\selectfont
\ASYalign(-218.645310,51.549640)(-1.000000,-0.500000){2}%
\definecolor{ASYcolor}{gray}{0.000000}\color{ASYcolor}
\fontsize{8.030000}{9.636000}\selectfont
\ASYalign(-218.645310,69.412514)(-1.000000,-0.500000){3}%
\definecolor{ASYcolor}{gray}{0.000000}\color{ASYcolor}
\fontsize{8.030000}{9.636000}\selectfont
\ASYalign(-218.645310,87.275389)(-1.000000,-0.500000){4}%
\definecolor{ASYcolor}{gray}{0.000000}\color{ASYcolor}
\fontsize{10.037500}{12.045000}\selectfont
\ASYalign(-219.247560,105.138263)(-1.000000,-0.320001){${\sim} W_c$}%
\definecolor{ASYcolor}{gray}{0.000000}\color{ASYcolor}
\fontsize{8.030000}{9.636000}\selectfont
\ASYalign(-218.645310,123.001138)(-1.000000,-0.500000){6}%
\definecolor{ASYcolor}{gray}{0.000000}\color{ASYcolor}
\fontsize{8.030000}{9.636000}\selectfont
\ASYalign(-218.645310,140.864012)(-1.000000,-0.500000){7}%
\definecolor{ASYcolor}{gray}{0.000000}\color{ASYcolor}
\fontsize{8.030000}{9.636000}\selectfont
\ASYalign(-218.645310,158.726886)(-1.000000,-0.500000){8}%
\definecolor{ASYcolor}{gray}{0.000000}\color{ASYcolor}
\fontsize{8.030000}{9.636000}\selectfont
\ASYalign(-218.645310,176.589761)(-1.000000,-0.500000){9}%
\definecolor{ASYcolor}{gray}{0.000000}\color{ASYcolor}
\fontsize{8.030000}{9.636000}\selectfont
\ASYalign(-218.645310,194.452635)(-1.000000,-0.500000){10}%
\definecolor{ASYcolor}{gray}{0.000000}\color{ASYcolor}
\fontsize{8.030000}{9.636000}\selectfont
\ASYalign(-46.825262,12.874029)(-0.500000,0.000000){0}%
\definecolor{ASYcolor}{gray}{0.000000}\color{ASYcolor}
\fontsize{8.030000}{9.636000}\selectfont
\ASYalign(-77.909858,12.874029)(-0.500000,0.000000){-10}%
\definecolor{ASYcolor}{gray}{0.000000}\color{ASYcolor}
\fontsize{8.030000}{9.636000}\selectfont
\ASYalign(-108.994454,12.874029)(-0.500000,0.000000){-20}%
\definecolor{ASYcolor}{gray}{0.000000}\color{ASYcolor}
\fontsize{8.030000}{9.636000}\selectfont
\ASYalign(-140.079050,12.874029)(-0.500000,0.000000){-30}%
\definecolor{ASYcolor}{gray}{0.000000}\color{ASYcolor}
\fontsize{8.030000}{9.636000}\selectfont
\ASYalign(-171.163646,12.874029)(-0.500000,0.000000){-40}%
\definecolor{ASYcolor}{gray}{0.000000}\color{ASYcolor}
\fontsize{8.030000}{9.636000}\selectfont
\ASYalign(-202.248242,12.874029)(-0.500000,0.000000){-50}%
\definecolor{ASYcolor}{gray}{0.000000}\color{ASYcolor}
\fontsize{10.037500}{12.045000}\selectfont
\ASYalign(-130.753671,5.999310)(-0.500000,-0.239646){$\log_2 \Xi$}%
\definecolor{ASYcolor}{gray}{0.000000}\color{ASYcolor}
\fontsize{8.030000}{9.636000}\selectfont
\ASYalign(-19.548585,24.981449)(0.000000,-0.500000){$10^{-5}$}%
\definecolor{ASYcolor}{gray}{0.000000}\color{ASYcolor}
\fontsize{8.030000}{9.636000}\selectfont
\ASYalign(-19.548585,70.079507)(0.000000,-0.500000){$10^{-4}$}%
\definecolor{ASYcolor}{gray}{0.000000}\color{ASYcolor}
\fontsize{8.030000}{9.636000}\selectfont
\ASYalign(-19.548585,115.177565)(0.000000,-0.500000){$10^{-3}$}%
\definecolor{ASYcolor}{gray}{0.000000}\color{ASYcolor}
\fontsize{8.030000}{9.636000}\selectfont
\ASYalign(-19.548585,160.275623)(0.000000,-0.500000){$10^{-2}$}%
}

  \caption {This density plot shows that the distributions of the logarithms of all of the elements of $\XXi$ collected together tend to be roughly bell shaped, with increasing disorder causing them to shift to the left.  The low values in the strongly disordered, localized cases reveal the failure of a large number of pairs of diagonal elements to significantly repel during the course of the flow, corresponding to the failure of Hilbert space percolation.  The horizontal line indicates the approximate location of the MBL transition, $W_c \approx 5$.\cite {RademakerOrtuno16}  The transition is not sharp due to the strong finite-size effects of a ten-site chain.  The curved line shows the medians of the distributions.  This data was averaged over four disorder realizations.}
  \hrule
  \label {DensityPlot}
\end {figure}

We now consider the distribution of the values of $\log {\Xi}$, treating all $ \left( n^2 - n \right) /2$ pairs of off-diagonal elements as a single statistical population.  As shown in figure~\ref {DensityPlot}, their distributions are roughly bell shaped.  Increasing the disorder strength spreads the distribution to the left.  For the strongly disordered cases, it is apparent that a large fraction of the pairs of diagonal elements repelled only trivially during the course of the flow.  This is compatible with the Poissonian level spacing expected in the many-body localized phase.  Due to the strong finite-size effects at play in this $L = 10$ system, it is unsurprising that the phase transition is not sharp.  We suspect that the nonmonotonicity in the low-disorder behavior is due to the almost preserved translation-invariance causing near-degeneracies, highlighting the integrability revealed by the Bethe ansatz.  One could consider also tracking the distance- or energy-dependence of the level repulsion behavior.

\section {Conclusion}

\subsection {The Interpretation of Continuous Unitary Transformations}

Traditionally, we think of CUTs as an inefficient way to diagonalize a matrix.\cite {MolerLoan03}  Instead, they occupy a similar conceptual niche as the strong-disorder renormalization group\cite {DasguptaMa79, DasguptaMa80, RefaelAltman13}.  Furthermore, we think of their true utility as providing an intuitive picture of the diagonalization of a Hamiltonian: specifically, a causal reinterpretation of perturbation theory.  Starting with an unperturbed Hamiltonian in its diagonal eigenbasis, adding a nontrivial perturbation generates off-diagonal elements.  If we then apply a flow to the perturbed Hamiltonian, we can watch as these off-diagonal elements decay in magnitude while repelling the diagonal elements they connect and rotating other off-diagonal elements into each other.  While truly small perturbations decay before disturbing the Hamiltonian too significantly, nonperturbative phenomena can be schematically visualized as ultimately having significant, global consequences over the whole of the matrix. 

Several flow generators have been suggested so far.  As discussed in section~\ref {Renormalization}, some choices for the infinitesimal flow generator, including the original WWF, can be approximately understood in terms of a renormalization group.\cite {GlazekWilson93, GlazekWilson94}  For example, the WWF preferentially tries to flow away off-diagonal elements which connect levels that already have a large energy difference, $\delta$.  It can therefore be thought of as flowing from large energy scales to small ones.

\subsection {Summary}

In section~\ref {NovelFlow}, we introduced the uniform tangent decay flow.  It is quite similar to the original WWF, with the key difference that it does not have this bias towards handling high-energy couplings more quickly.  Whether this is beneficial depends on the application.  Among the advantages of the tangent flow are that it does not exhibit the stiffness of the WWF:  off-diagonal elements decay in synchrony, thereby increasing the efficiency of unstabilized integrators such as the Dormand--Prince method, and it is invariant under energy rescalings.  In contrast, halving the energy scale of the WWF quadruples the fictitious flow-time scale.  However, the tangent flow's small-radius off-diagonal elements are quite sensitive to the motions of the larger-radius elements, and this, in turn, can slow down numerical integration.

On top of the analytic value of CUTs, they also constitute an important numerical technique.\cite{PekkerEtAl16, BenediktUhrig13, KronesUhrig15}  Thus, it is important that we have efficient algorithms available for the implementation of CUTs \emph {in silico}.  However, the existing literature on this topic could be substantially expanded.  With this motivation in mind, we have improved upon the existing integrators such as the Dormand--Prince method by presenting three integrators designed specifically for the purpose of CUTs.

In most situations, we recommend taking advantage of the stability, unitary exactness, and third-order convergence properties of the integrator developed in section~\ref {ThirdOrder}.  It should not be difficult to extend it to flows other than the WWF and the tangent flow, so long as the derivatives of the generator can be calculated and the ansatz in equation~\eqref {Integral} is modified appropriately.  The primary advantage of its first-order predecessor in section~\ref {FirstOrder} is its relative simplicity, although its stability might also be more robust for some flow generators.  The Trotter integrator in section~\ref {Trotter} does not appear to be immediately optimal for any of our concrete applications, but has intriguing conceptual, stability, and computational complexity properties, and so was included for the sake of completeness.

We took advantage of the third-order integrator and applied it to an MBL problem in section~\ref {MBL} in order to explore whether flows provide insight into the one-dimensional MBL transition.  In particular, we sought to track the spread of the off-diagonal couplings through the Hamiltonian matrix in order to better understand the asymptotic scaling behavior of a perturbation at its critical intensity.  The analysis revealed a clear connection between the distribution of integrated level repulsion strengths and the ergodicity breaking transition. It also captures the integrability via the Bethe ansatz of the model at zero disorder.

\subsection {Future Directions}

For future work, we suggest considering whether the unperturbed Hamiltonian provides a natural sense of ``distance'' between pairs of its eigenstates.  If so, one can watch as the presumably initially short-range perturbation ``expands'' into longer-range interactions during the course of the flow.  Phase transitions, including those involving the debated nonergodic extended phase\cite {SerbynMoore16, BiroliEtAl12}, should correspond to qualitative changes in the competition between the rate of this interaction expansion, the decay of the off-diagonal elements, and the exponential growth of the number of sites at a given distance.  One signature of the nonergodic extended phase should be $\Xi$ distributions at long distances with exponentially small fractions of --- but still exponentially many --- pairs above some arbitrary repulsion threshold.  We hope to apply this analysis to the paradigmatic models of MBL consisting of Anderson localization on a high-dimensional\cite {TarquiniEtAl17} or hyperbolic space, such as a $2^N$ site hypercube\cite {AltshulerEtAl10} or the Bethe lattice\cite {AndersonEtAl73, AltshulerEtAl16}.

Second, all of the stable fixed points of the flows listed in table~\ref {Flows} are diagonal.  As mentioned in section~\ref {Introduction}, the resulting permutation of the eigenvectors can be used to construct local integrals of motion for localized Hamiltonians.\cite {QuitoEtAl16}\textsuperscript ,\cite {RademakerOrtuno16}  Perhaps flows with a larger set of stable fixed points would provide even more insight into the relevant physics.  In particular, we are considering changing the objective function in equation~\eqref {SteepestDescent}:  For the WWF, it is $j^2$, which we suggest modifying to $x^2 j^2$.  The resulting generator changes from $\eta = r^2 \sin {2 \theta} = \delta j$ to $r^4 \sin {4 \theta} = 4 x j (x^2 - j^2)$.  The set of stable fixed points is now characterized by block diagonal matrices, possibly permuted, with constant diagonals within each block.  The stabilizing decay ansatz employed in the construction of the third-order integrator in section~\ref {ThirdOrderStabilization} should not be difficult to generalize to this flow. 

A final potential avenue for future research involves constructing the generator $\EEta$ out of something other than just the Hamiltonian $\HH$.  In particular, we imagine that a wavefunction $\psi$ co-evolving according to both a CUT flow and Schr\"odinger evolution, $\dot \psi = (\EEta - i \HH)\psi$, could provide useful information for flows designed to highlight localization phenomena.

\begin {acknowledgments}
    This work was supported by the Institute of Quantum Information and Matter, a National Science Foundation frontier center partially funded by the Gordon and Betty Moore Foundation.  G.~R.~acknowledges the generous support of the Packard Foundation and the National Science Foundation through award DMR-1410435.  Thanks to Evert van Nieuwenburg, Stefan Kehrein, Marcus Bintz, Christopher White, Alex Bourzutschky, and Paraj Titum for many fruitful discussions.  The numerical CUT flows were implemented using double-precision floating-point matrices calculated by the open-source linear algebra library \textsc {Armadillo}\cite{SandersonCurtin16}.  Special thanks to Changnan Peng for GPU-accelerating the flows in section~\ref {UnitaryIntegrator} using the open-source linear algebra library \textsc {ArrayFire}\cite {YalamanchiliEtAl15} and thereby helping to test their performance. 
\end {acknowledgments}

\bibliography {library}

\end {document}